\documentclass{article}

\usepackage{mathtools}
\usepackage[mathcal]{euscript}
\usepackage{amssymb}
\usepackage{xcolor}
\usepackage{bm}
\usepackage{subcaption}
\usepackage{setspace}

\usepackage[margin=2cm]{geometry}
\usepackage{newtxtext}
\usepackage{newtxmath}

\newcommand{\z}[1]{\textcolor{black}{#1}}
\newcommand{\y}[1]{\textcolor{black}{#1}}

\newcommand{\x}[1]{\textcolor{black}{#1}}

\DeclarePairedDelimiter{\norm}{\lVert}{\rVert}

\usepackage{tabstackengine}
\setstacktabbedgap{1.5ex}               
\setstackgap{L}{1.2\normalbaselineskip}

\usepackage[absolute]{textpos}
\usepackage{xcolor}
\begin{document}

\begin{textblock}{17}(3.5,1)
	\noindent\textbf{\color{red}Published in ``IEEE Robotics and Automation Magazine'', DOI: 10.1109/MRA.2020.3044911}
\end{textblock}

\title{Astrobotics: Swarm Robotics for Astrophysical Studies}               

\author{Matin~Macktoobian, Denis~Gillet, and Jean-Paul~Kneib%
	\date{This work was financially supported by the Swiss National Science Foundation (SNF) Grant No. 20FL21\_185771 and the SLOAN ARC/EPFL Agreement No. SSP523. \textit{(Corresponding author: Matin Macktoobian.)}}%
	\date{The authors are with the School of Engineering, Swiss Federal Institute of Technology in Lausanne (EPFL), Lausanne, Switzerland (e-mail: matin.macktoobian@epfl.ch; denis.gillet@epfl.ch; jean-paul.kneib@epfl.ch).}%
}

\maketitle

\begin{abstract}
This paper introduces the emerging field of astrobotics, that is, a recently-established branch of robotics to be of service to astrophysics and observational astronomy. We first describe a modern requirement of \z{dark matter studies}, i.e., the generation of the map of the observable universe, using astrobots. Astrobots differ from conventional two-degree-of-freedom robotic manipulators in two respects. First, the dense formation of astrobots give rise to the extremely overlapping dynamics of neighboring astrobots which make them severely subject to collisions. Second, the structure of astrobots and their mechanical specifications are specialized due to the embedded optical fibers passed through them. We focus on the coordination problem of astrobots whose solutions shall be collision-free, fast \x{execution}, and complete in terms of the astrobots’ convergence rates. We also illustrate the significant impact of astrobots assignments to observational targets on the quality of coordination solutions To present the current state of the field, we elaborate \x{the open problems including} next-generation astrophysical projects including $\sim$20,000 astrobots\z{, and other fields, such as space debris tracking, in which astrobots may be potentially used}.
\end{abstract}
\doublespacing
\section{From Observation Definition to Survey Generation}
\begin{figure}
	\centering
	\includegraphics[scale=1]{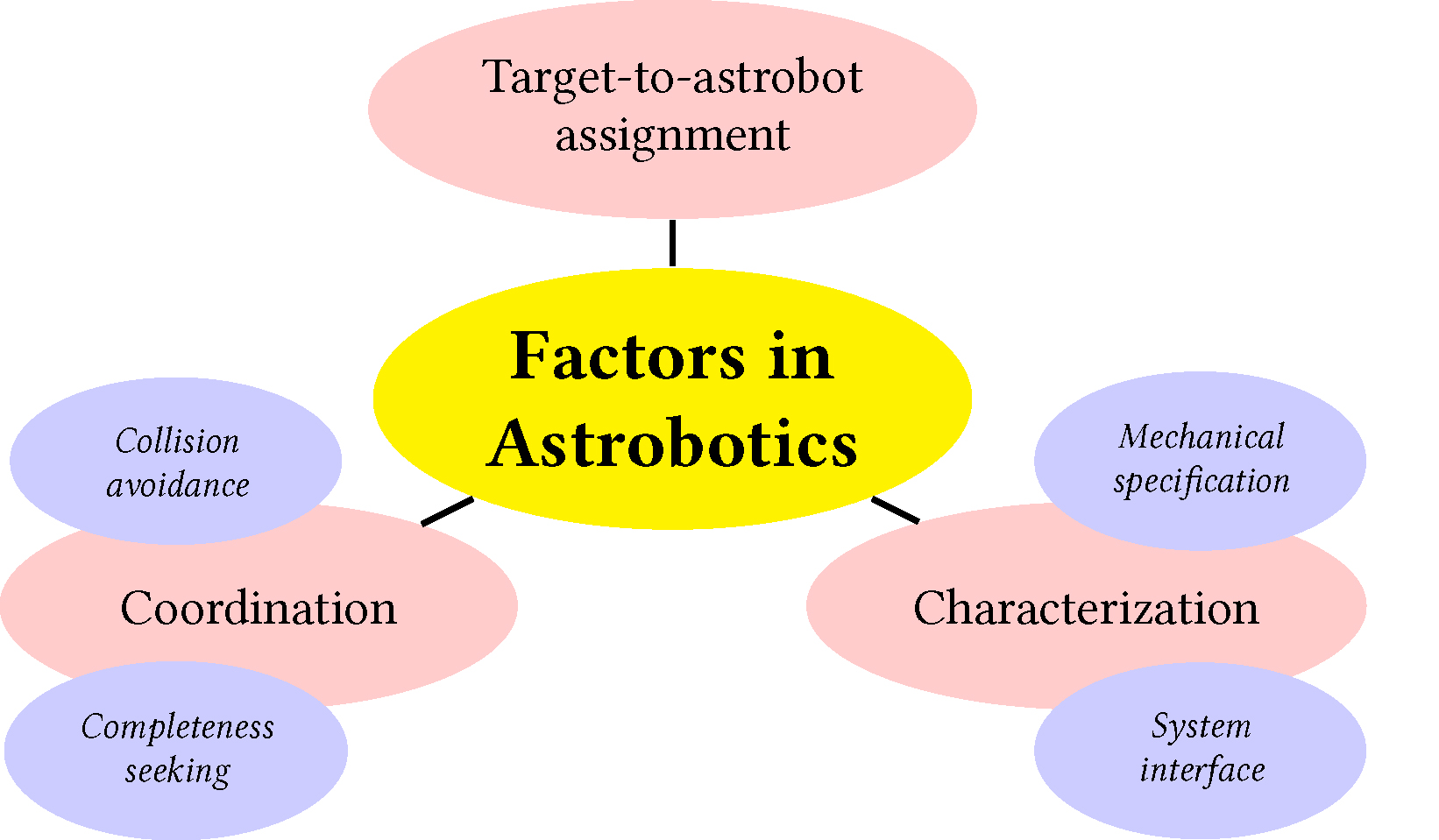}
	\caption{Factors in Astrobotics}
	\label{fig:fac}
\end{figure}

\begin{figure}
	\centering
	\includegraphics[scale=0.25]{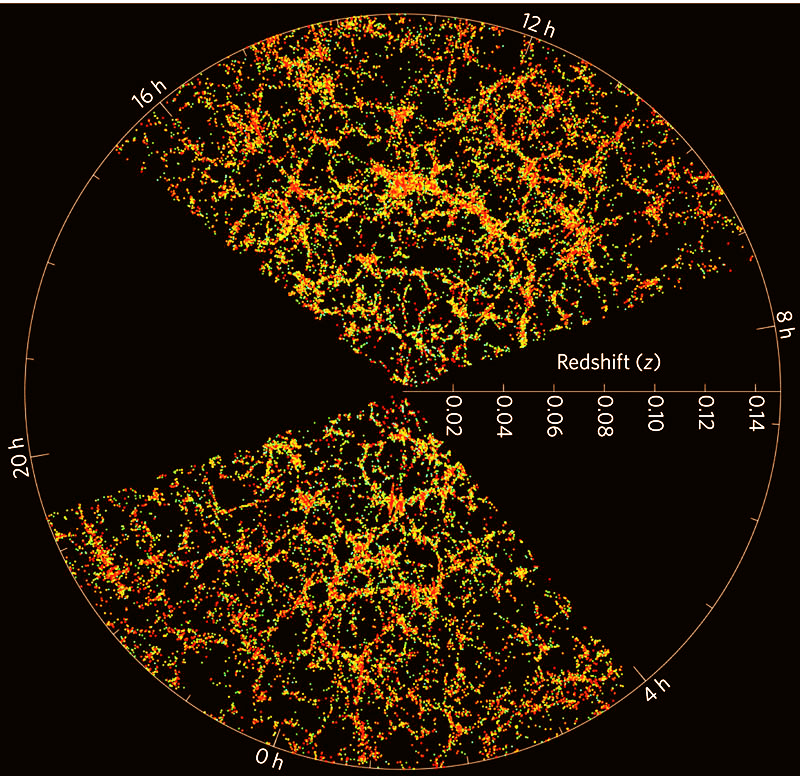}
	\caption{A survey representing the distribution of a cluster of galaxies obtained from the SDSS project (the image reprinted from SDSS website)}
	\label{fig:dist}
\end{figure}

Astrobotics is an emerging field of swarm robotics aiming to the development and control of astrobots \cite{horler2018robotic,horler2018high} to be of service to astrophysical studies and cosmological spectroscopic observations. In particular, astrobotics addresses a wide range of swarm-robotic-related topics (see, Figure \ref{fig:fac}) which exhibit challenging problems in design, interaction, coordination, and mission planning corresponding to astrobots. There have been many astrophysical projects, such as \x{the} SDSS family \cite{catinella2012galex} which seek the generation of the map of the observable universe. Such a map is known to open new venues to study the nature and the distribution of dark \z{matter} \x{throughout} the universe. To this aim, astrobots have already shown deep merit in \x{automation} of the cosmological observations. Astrobots not only improve the efficiency of observations but also provide more data of the sky to generate better maps with higher resolutions. To generate the map of the observable universe, a recent technology, called cosmological spectroscopy, is established. The sky is divided into \x{observational sectors, each of which includes} a subset of the intended astrophysical objects, such as quasars, galaxies, etc. Then, a ground telescope has to be equipped with the required spectroscopic utilities. In particular, a set of optical fibers needs to be mounted on a special area of the telescope which is known as focal plane. Then considering an observation, the idea is the assignment of each target of the observation to one of the fibers. In the course of the exposure time of an observation, each fiber collects the visible light \x{emitted} from its target. The light signals of all fibers are then transferred to a signal processing \x{device} called spectrograph (see, Figure \ref{fig:spec}) whose output is a spectroscopic survey (see, Figure \ref{fig:dist}). Eventually in the course of performing thousands of observations, the combination of all of the obtained surveys gives rise to the map of the observable universe. The careful analysis of those surveys and the final map is expected to reveal new discoveries about dark matter spread over the cosmos. Since the locations of the targets corresponding to each observation differs those of the other observations, one has to change the configuration of fibers so that each one of them points to its new target. This coordination process had been manually done by operators in the first generations of the cited surveys. However, the quoted manual coordination is indeed disadvantageous because it is cumbersome knowing that the number of fibers may exceeds thousands. Second, each observation has to be performed in a particular time \x{window}. Additionally, the available time is limited to coordinate astrobots between two consecutive observations. Thus, if an observation includes a massive number of targets, the whole fiber set may not be prepared on time before the observation time. Thus, the idea is \x{pass} each fiber through an astrobot which can automatically coordinate its corresponding fiber. \z{It is worth noting that the idea of distributed signal collection from \x{the} sky \x{has} been also taken into account
in multiple-mirror telescopes \cite{hege1985multiple}. In these telescopes, small mirrors are first calibrated by bright \x{stars} then coordinated to point to their targets. Compared to optical fibers and astrobots, these mirror-based systems are used for interferometric purposes. Furthermore, the density of their placements is \x{lower than that of} optical fibers. So, their coordination \x{is} less crucial in view of potential collisions.}

Given an observation and an astrobots swarm, one needs a mapping procedure to assign each target to one of the available astrobots. The smaller the astrobots are, the more astrobots one can mount on a focal plane, so the more targets an observation can contain. However, the coordination becomes more challenging, as well. In particular, astrobots are placed in a hexagonal formation next to each other, and their working spaces overlap those of their neighboring peers \z{(see, Figures \ref{fig:swarm}-\ref{fig:overlap})}, that is, the system is inherently prone to collisions. Thus, any control solution must resolve all potential collisions. Another coordination-related issue is the completeness. The coordination problem of astrobots is complicated due to their dense and highly-overlapping placement. Thus, a coordination solution may not be able to preserve both the safety and the complete coordination of a set of astrobots. The more astrobots are coordinated, the more information of an observation are obtained, thereby the more complete the final survey is. So, any safe coordination strategy which also fulfills the stated completeness is of utmost interest.

Thanks to the availability of huge ground telescopes with large focal planes, the recent and the future surveys are planned to host thousands of fibers as reported in Figure \ref{fig:number}. For example, it is practically impossible to manually coordinate all 20,000 multiplexed fibers of the MegaMapper instrument \cite{schlegel2019astro2020} considering the fairly limited amount of time available between consecutive observations for coordination. That is why the \x{automation} of coordination using astrobots is extremely demanding and necessary.
\section{Astrobot Characterization}
\begin{figure*}
	\centering
	\hspace*{-15mm}
	\begin{subfigure}[b]{.15\textwidth}
		\centering
		\includegraphics[scale=1]{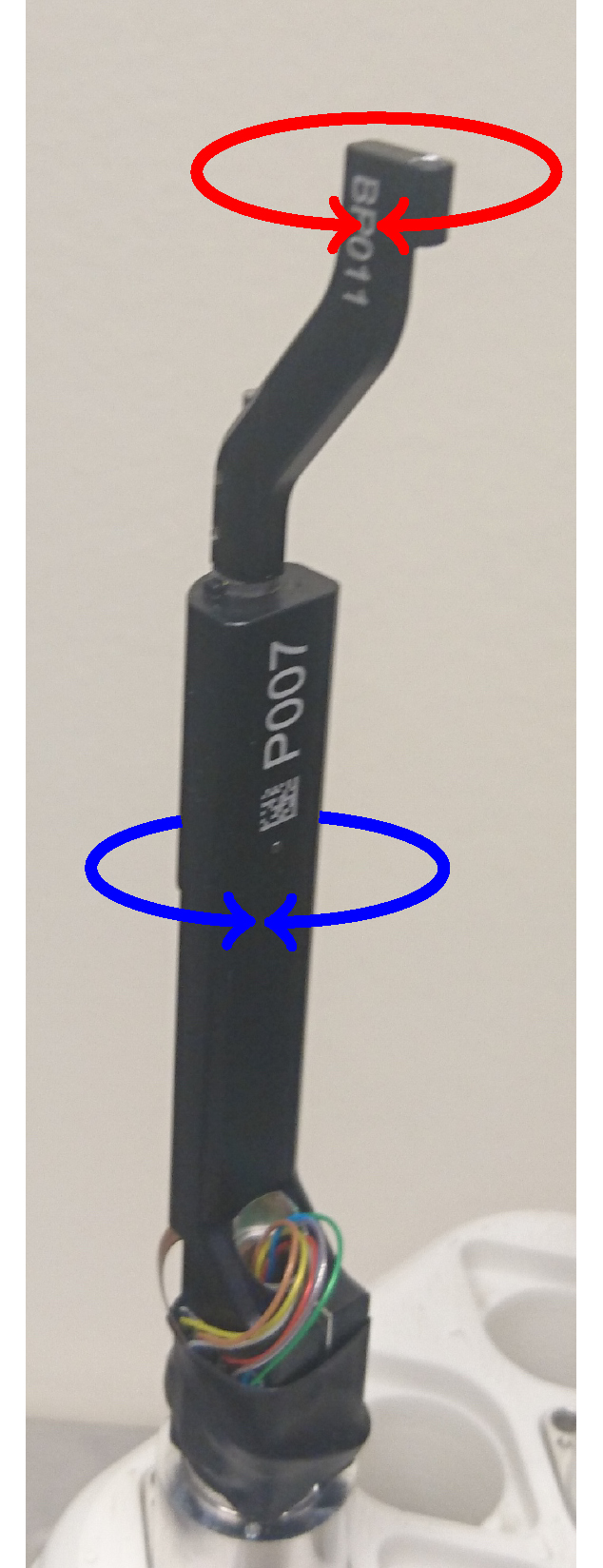}
		\caption{\label{fig:xxx}}
		\vspace{-7mm}
	\end{subfigure}
	\begin{subfigure}[b]{.45\textwidth}
		\centering
		\includegraphics[width=1.05\textwidth]{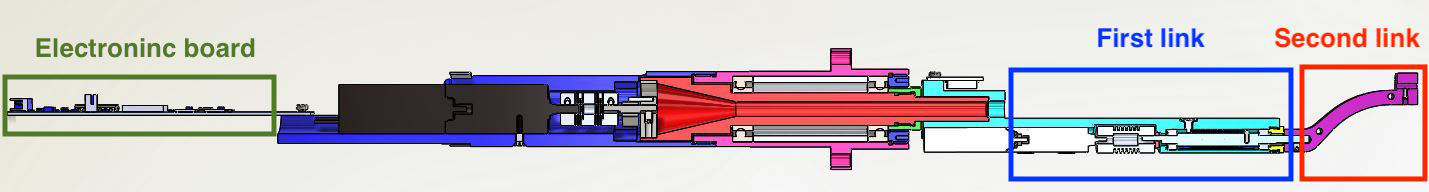}
		\caption{\label{fig:pos}}%
		\vspace{-2mm}	
		\begin{minipage}[b]{.5\linewidth}
			\centering
			\subcaptionbox{\label{fig:top}}
			{\includegraphics[scale=0.4]{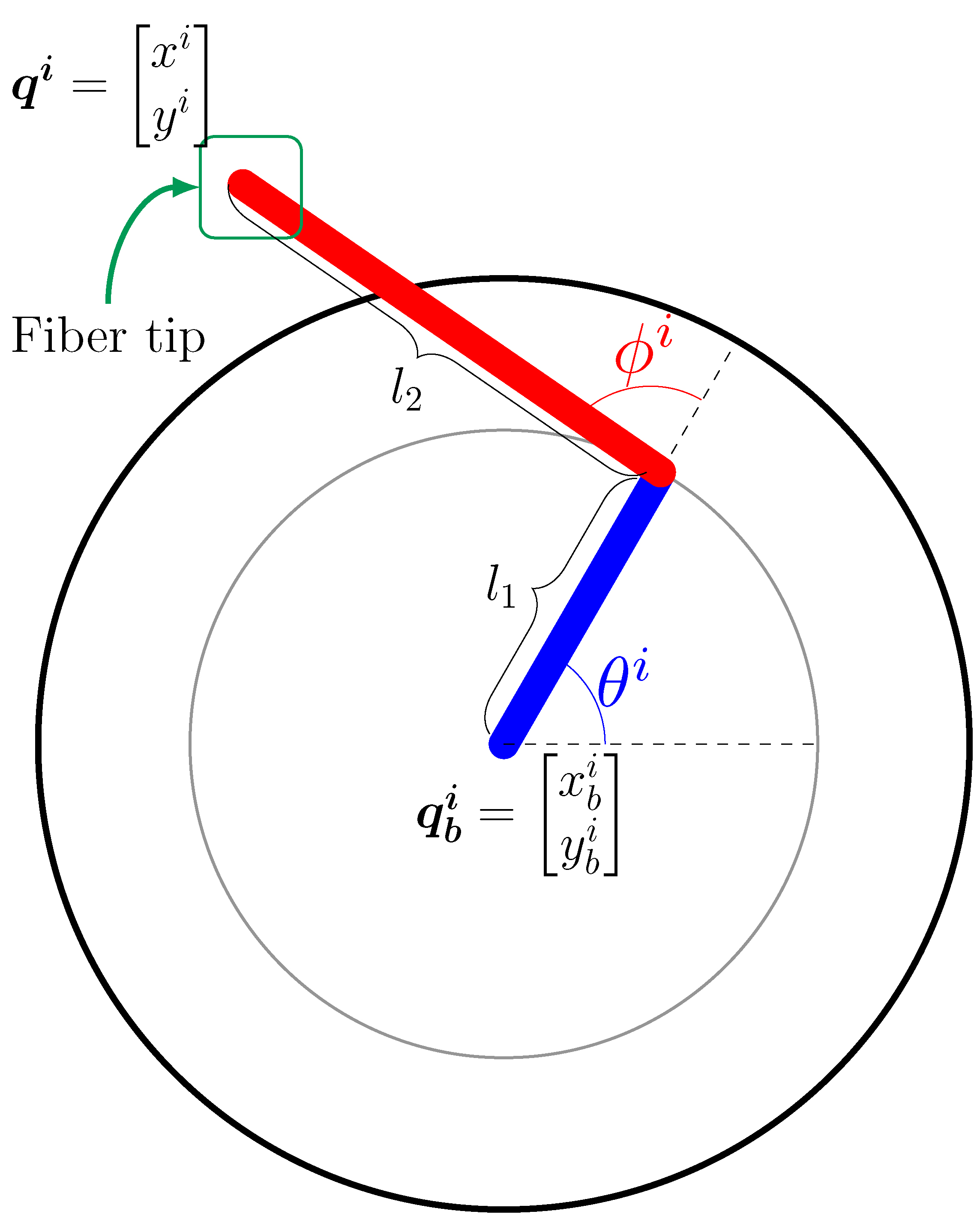}}
		\end{minipage}\quad
		\begin{minipage}[b]{.4\linewidth}
			\centering
			\subcaptionbox{\label{fig:spec}}
			{\includegraphics[scale=0.068]{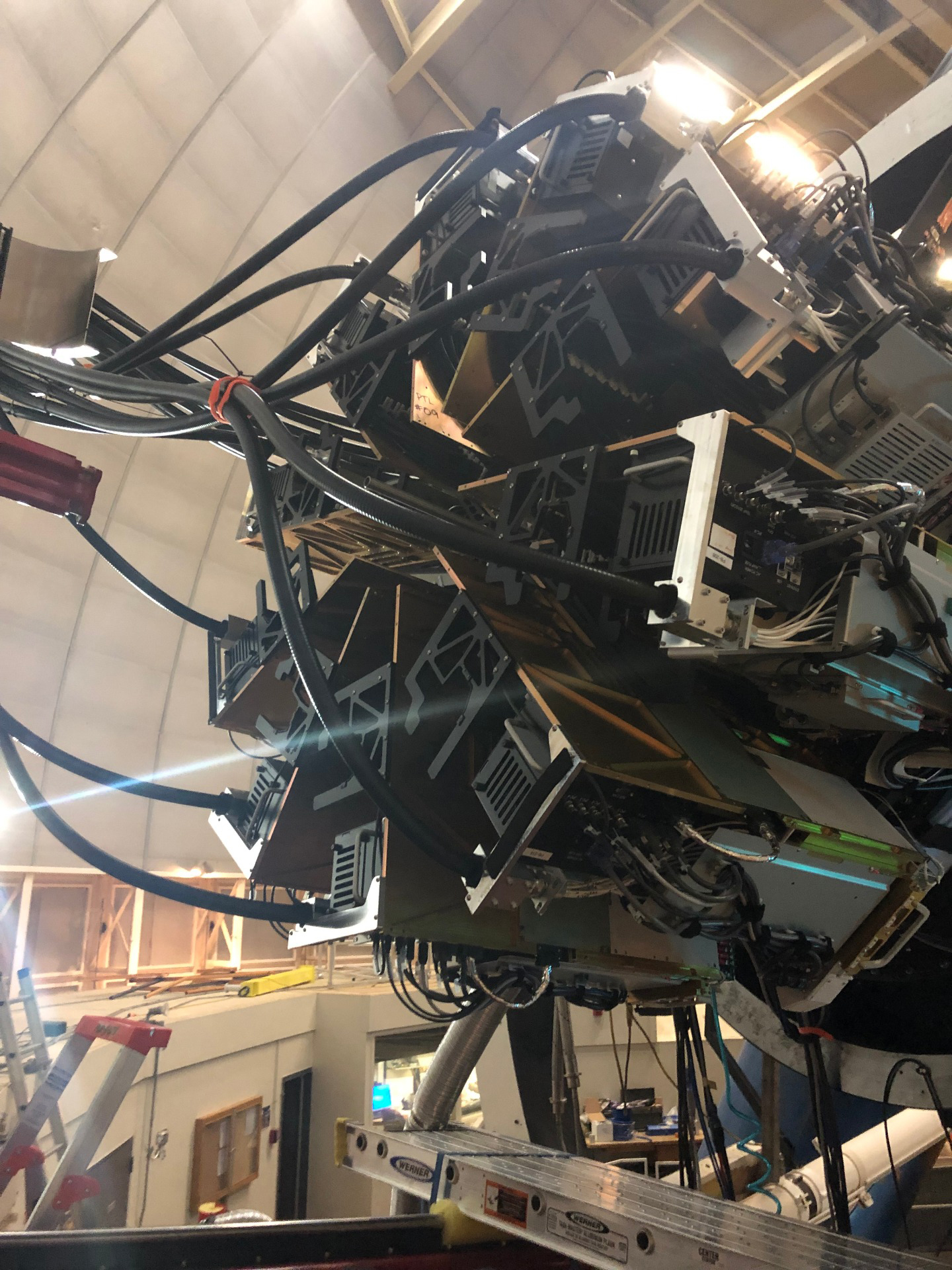}}
		\end{minipage}
	\end{subfigure}
	\begin{subfigure}[b]{.3\textwidth}
		\centering
		\subcaptionbox{\label{fig:swarm}}
		{\includegraphics[scale=0.65]{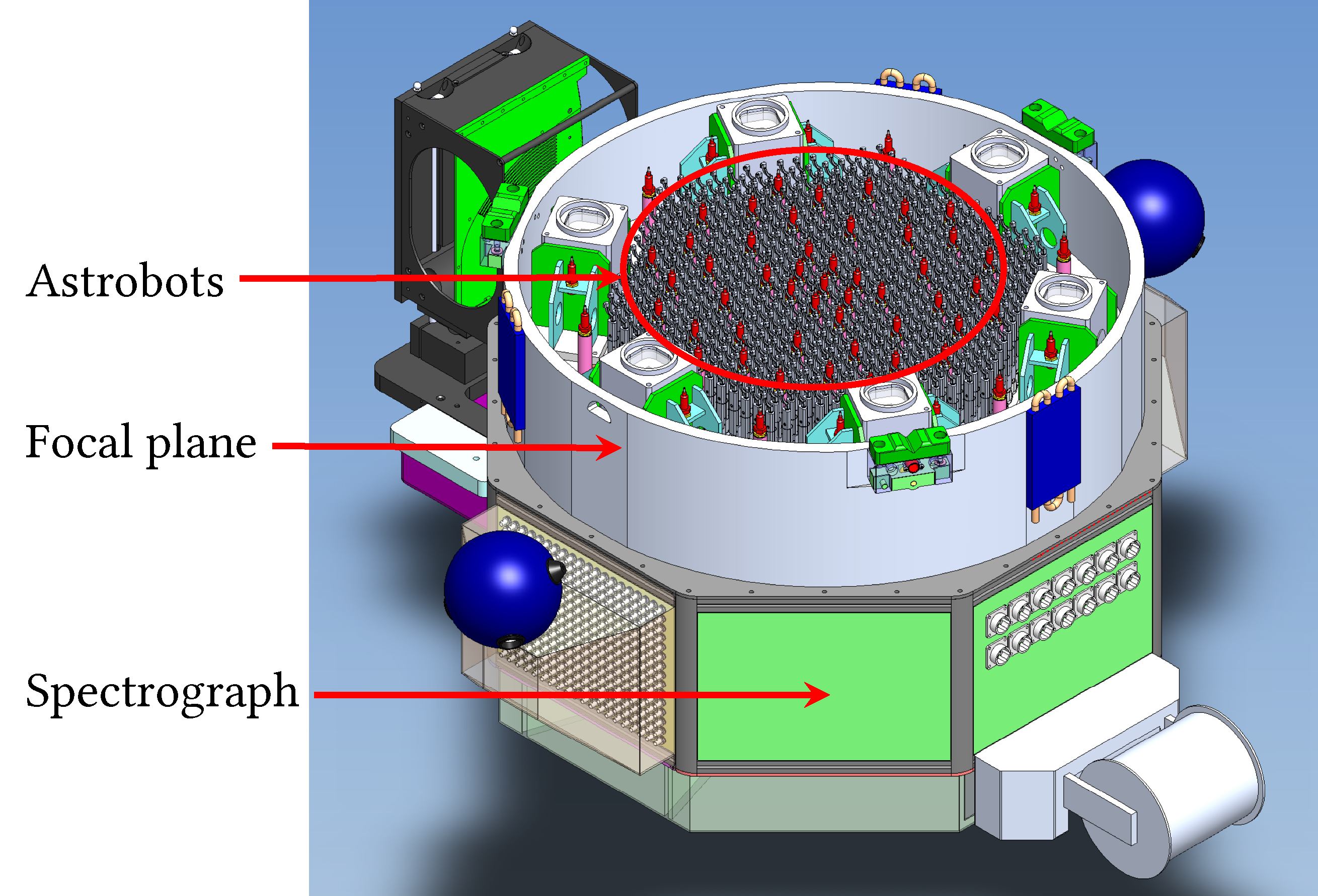}}
	\end{subfigure}
	\begin{minipage}[b]{.4\linewidth}
		\centering
		\subcaptionbox{\label{fig:overlap}}
		{\includegraphics[scale=0.9]{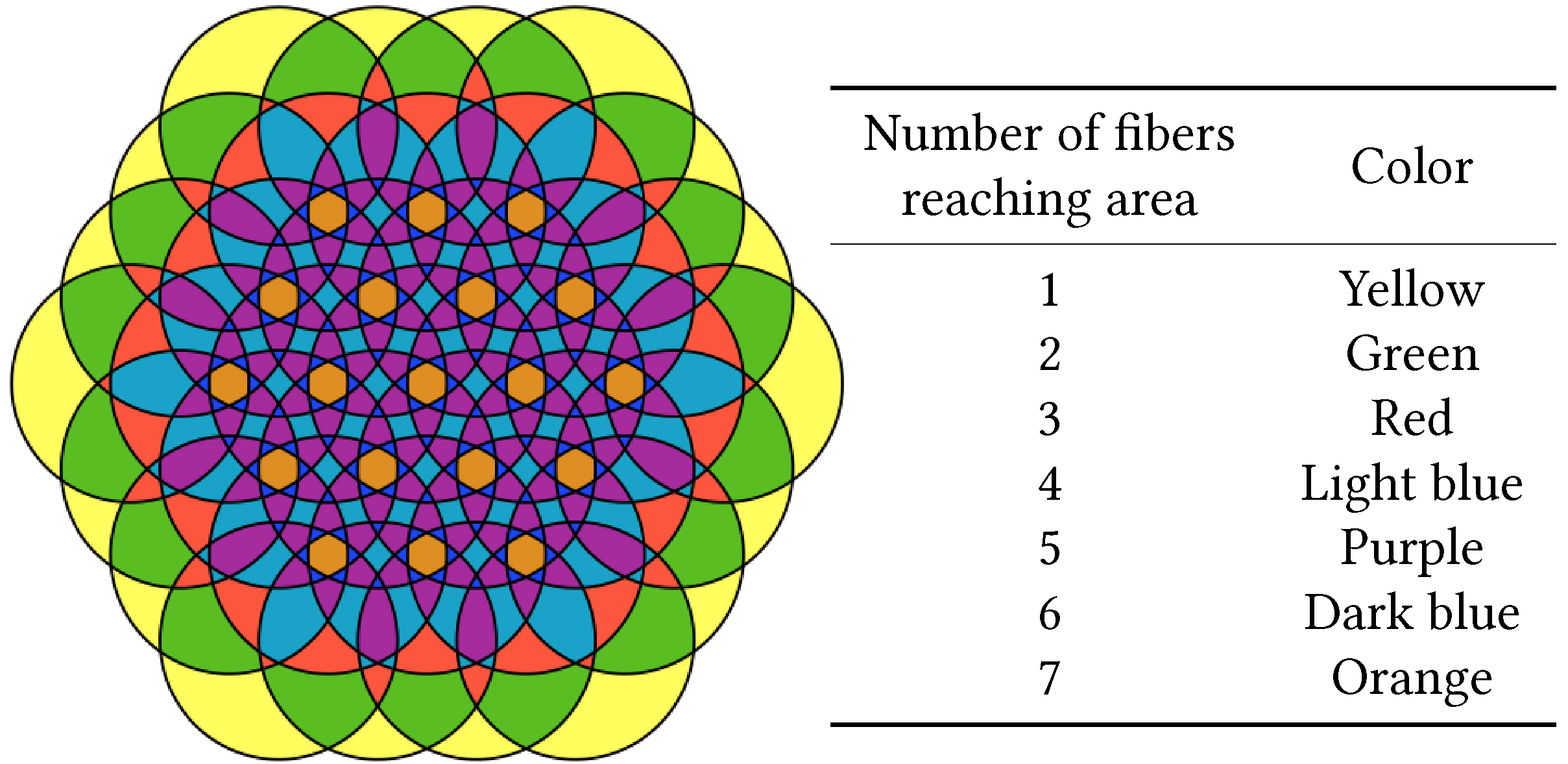}}
	\end{minipage}
	\hfill
	\begin{minipage}[b]{.45\linewidth}
		\centering
		\subcaptionbox{\label{fig:number}}
		{\includegraphics[scale=0.85]{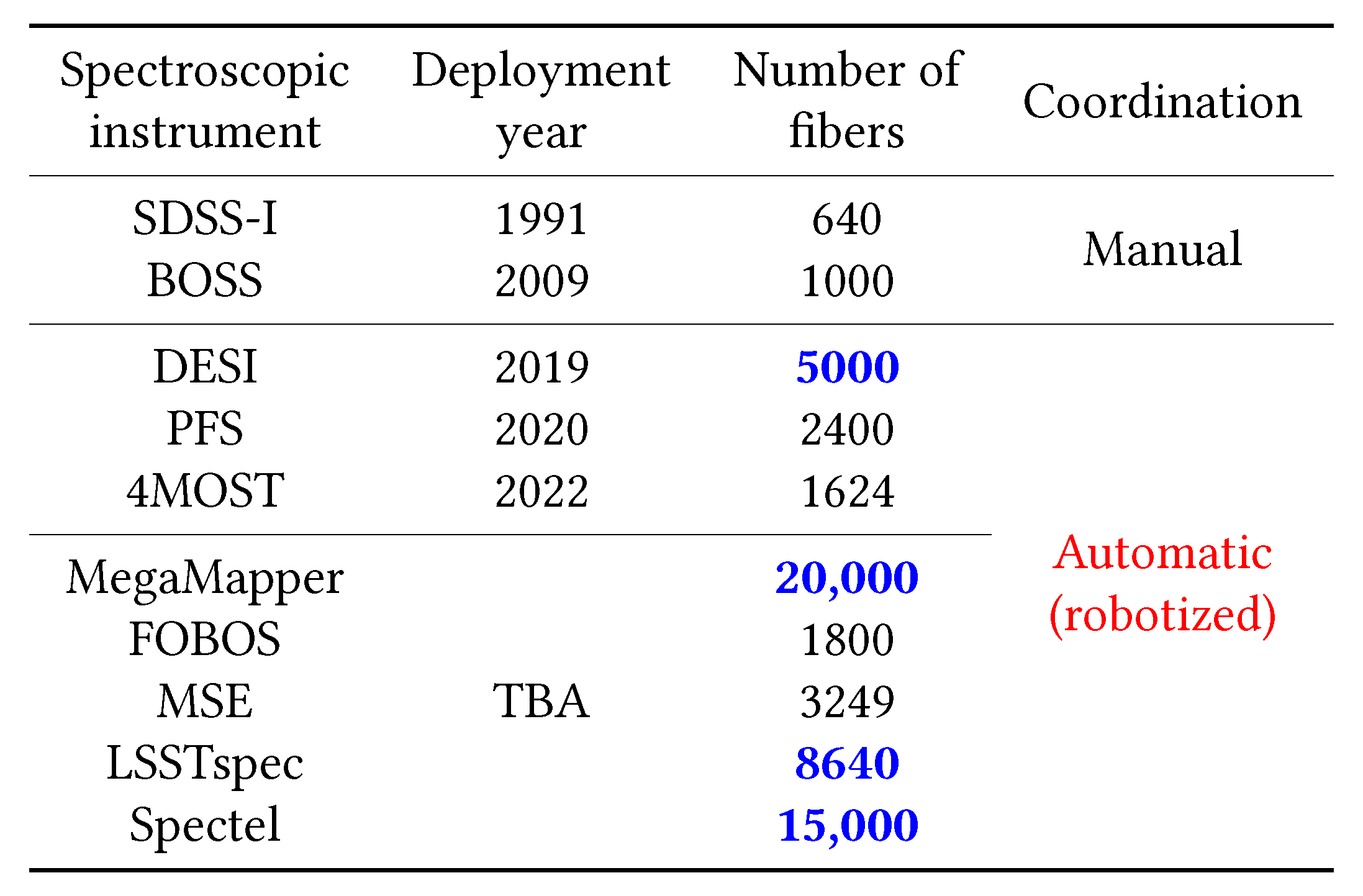}}
	\end{minipage}
	\caption{\z{Astrobots swarm system (subfigures b and f are partially reprinted from \cite{horler2018robotic}; subfigures c and e are reprinted from \cite{macktoobian2019navigation} and DESI webpage in Ohio State University, respectively) (a) 3D representation of the rotational degrees of freedom of an astrobot. (b) The side view of an astrobot. (c) The top view of an astrobot whose parameters indicate the kinematics of its ferrule's motion. (d) A spectrograph which processes the signals collected by astrobots to generate a survey. (e) An astrobots swarm located at a telescope's focal plane. (f) Overlapping pattern of astrobots in a neighborhood illustrating their collision-prone motions. (g) The plan of the past, current, and future spectroscopic surveys (the data are adapted from \cite{ferraro2019inflation})}}
\end{figure*}

Each astrobot is a SCARA-like two-degree-of-freedom rotational-rotational robotic manipulator as depicted in Figures \ref{fig:xxx} and \ref{fig:pos}. An optical fiber is mounted onto the astrobot such that its tip is located at the end-effector of the astrobot, called ferrule. Rotational combinations of the astrobot's arms move the ferrule in the working space of the astrobot such that it can reach some targets corresponding to a planned observation. Astrobots are placed in a honeycomb-like hexagonal formation on focal planes as illustrated in Figure \ref{fig:swarm}. Given\footnote{Throughout this paper, scalars and matrices are represented by regular and bold symbols, respectively.} an astrobot $i$ associated with a swarm, let $\bm{q^{i}_{b}}= \bracketMatrixstack{x^{i}_{b} & y^{i}_{b}}^\intercal$ be the coordinate of the astrobot's base. This coordinate can be defined as a part of a universal frame attached to the focal plane of its host telescope. The lengths of the links are denoted by $\bm{l} = \bracketMatrixstack{l_{1} &l_{2}}^\intercal$. Then, the location of the ferrule, say, $\bm{q^{i}} = \bracketMatrixstack{x^{i} & y^{i}}^\intercal$, is stated as below given the angular deviations of the astrobot's arms represented by $\theta^{i}$ and $\phi^{i}$ (see, Figure \ref{fig:top}). 
\begin{equation}\label{eq:kin}
\bm{q^{i}} = \bm{q^{i}_{b}} + 
\bracketMatrixstack{\cos(\theta^{i}) & \cos(\theta^{i} + \phi^{i})\\
	\sin(\theta^{i}) & \sin(\theta^{i} + \phi^{i})} \bm{l}
\end{equation}
\section{Challenges and Solutions}
\begin{figure}
	\centering\includegraphics[scale=0.6]{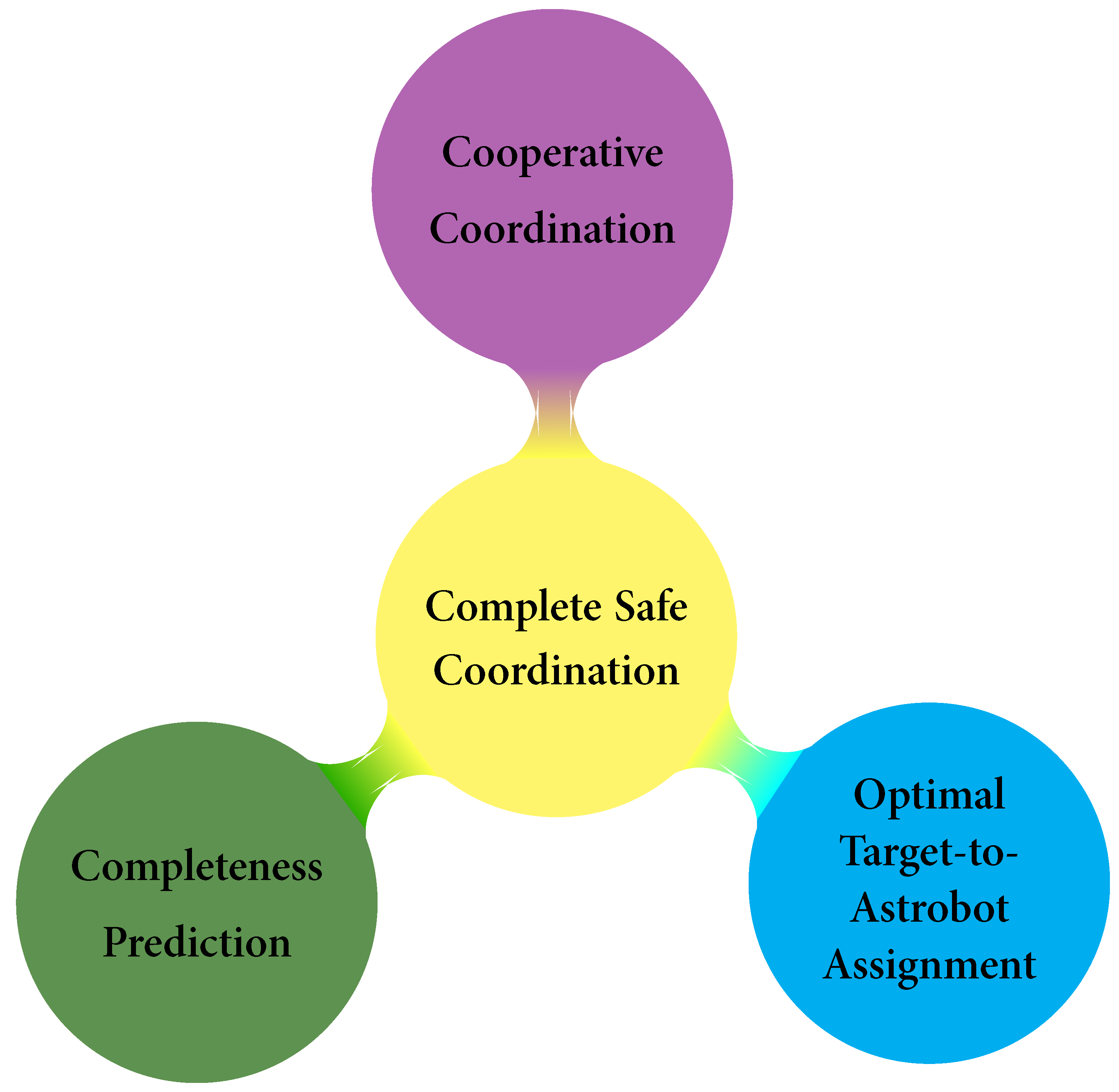}
	\caption{\y{The elements of the solution to the complete safe coordination problem}\label{fig:mindmap}}
\end{figure}

\y{The astrobots must avoid any collisions with their neighboring peers. Moreover, all of the astrobots of a focal plane need to reach their targets to maximize the information throughput of each observation. Thus, the main challenge is the safe complete coordination of \x{astrobot} swarms. This problem may be partitioned to three sub-problems, discussed in the following sections, whose solutions contribute to the desired coordination. In particular, observation planning in terms of assigning targets to astrobots significantly influence coordination. One seeks the assignments which minimize coordination times and maximize the distributions of astrobots in a focal plane. Second, completely independent coordination of each astrobot may lead to many deadlocks in the course of coordination because each astrobot is indifferent to the convergence of its peers. So, control synthesis for full convergence may realize completeness. Finally, completeness checking requires intensive time- and resource-consuming simulations. One may be interested in testing those coordination scenarios which potentially reach a minimum convergence rate. Put differently, the data associated with previous coordination results of a swarm may be used to synthesize a machine-learning-based convergence predictor. Thus, given an observation and its assigned pairs of astrobots-targets, the predictor may estimate the overall convergence of the swarm without any simulation. If the result is lower than \x{the} minimum, then the assignment may be revised without \x{the need for further} simulations. In the following sections, we describe the details of these problems and their current solutions conceptually, as depicted in Figure \ref{fig:mindmap}. We also \x{discuss} future research trends in this field to \x{guide} the research by a broader robotics community.}
\subsection{Target-to-Astrobot Assignment}
\x{An individual observation consists of the detection of light of a set of astrophysical objects over a finite period of time. Galactic catalogs that list object names and coordinates are typically used to make assignments of targets within an observation} \cite{anderson2014clustering}. \x{Since we are primarily interested in the coordination of the astrobots, we assume that we are working with a set of known targets that are ready to be assigned to astrobots}. Each target of an observation needs to be assigned to one of the available astrobots whose ferrule can reach the location of the target projected onto the focal plane. Since each target may be reachable by many astrobots in a locality, the key question is how to address the assignment process. 

The idea of random assignment assigns each target to a random un-allocated astrobot which reaches it. \x{This method is complicated by the lack of performance criteria used for the assignment}. One may note that random assignment may construct a dense locality of astrobots assigned to some targets while other reaching astrobots are left in farther distances from the locality. In this case, the astrobots' coordination may not yield high convergence rates because of the deadlocks they may encounter in their dense neighborhoods. As another solution to the problem, the drainage algorithm \cite{morales2011fibre} moves targets among different lists of un-allocated astrobots which reach them. To ensure the maximum number of targets handled during an observation, the method assigns each target to the astrobot whose list is the shortest with respect to observable targets. This method also lacks any internal mechanism to take physical size of astrobots into account. Thus, one may not estimate the potential collisions which may rise during coordination associated with a particular assignment. Target-based assignment method \cite{schaefer2016target}, however flips the problem by assigning astrobots to targets to yield higher assignment ratios.

\y{The assignment process is extremely influential on the realization of safe, fast, and complete coordination. But none of the stated methods above effectively formulate and address those requirements in the course of assignments. In this regard, optimal target assignment \cite{macktoobian2020optimal} presents the first solution to the problem whose impact on the coordination phase is fully specified. Namely, this method performs assignment constrained to two performance and safety requirements. The minimum coordination requirement seeks the fastest coordination corresponding to an astrobot, say, a target is assigned to an astrobot whose initial ferrule's coordination is in the closest distance to the projection of the target on the host focal plane compared to the distance of that projection to other astrobots reaching the target. Moreover, safety and completeness are addressed using the maximum distribution requirement which is formulated as follows. A target has to be assigned to an astrobot which is the farthest astrobot, among those reaching the target, with respect to the already-assigned astrobots. This requirement seeks the maximum dispersion of assigned astrobots all over a focal plane, so that their interactions, i.e., deadlocks and/or collisions, are minimized.}

\y{One may note that the formulation of the optimality criterion proposed in \cite{macktoobian2020optimal} based on standard optimization problems makes the problem solving process very inefficient because both the cost function and the class of constraints will be extremely nonlinear. Then, it becomes unlikely that global optimal solutions may be efficiently obtained.}
\subsection{Coordination Control}
The configurations of optical fibers are unique with respect to any observation because each observation comprises different targets. Thus, a reconfiguration procedure is a must to move fibers from one observation to another. The manual reconfiguration of fibers were extensively taken into account in the first generations of spectroscopic surveys, e.g., SDSS I-IV \cite{dawson2012baryon}. In these projects, for each observation, a new plate had to be assembled so that fibers point to the targets of the observation. Then, the plate is located at the focal plane to perform a planned observation. Both the cost of multiple plates' manufacturing and the substantial labor to manually transfer astrobots from one plate to another one are not negligible. SDSS-V project, though, uses fully-automated focal planes whose astrobots have to be coordinated. \y{The cited coordination have to suffice both collision freeness and complete convergence requirements of astrobots.}
\subsubsection{Collision Avoidance}
Each coordination operation seeks a particular configuration of astrobots. Thus, one expects a coordination process stops only when its desired configuration is reached. For this purpose, distributed navigation functions (DNFs) \cite{de2006formation} inherently avoid collisions. \y{In safety-critical applications, such as astrobotics, one enjoys the intrinsic safety and fast convergence of the control synthesized by these class of coordinators compared to other strategies \cite{rossi2018review}.} In particular, each astrobot was assigned to a distributed navigation function as follows \cite{makarem2014collision}
\begin{equation}
\psi(\bm{q^{i}}) := \underbrace{\vphantom{\bm{\lambda_{2}}\displaystyle\sum\limits_{\mathclap{j \in \mathcal{N}^{i}\setminus \{i\}}}          \min [0,\frac{\norm{\bm{q^{i}} - \bm{q^{j}}}^{2} - D^{2}}{\norm{\bm{q^{i}} - \bm{q^{j}}}^{2} - d^{2}}}\bm{\lambda_{1}}\norm{\bm{q^{i}} - \bm{q^{i}_{\mathcal{T}}}}^{2}}_{\text{attractive term}} +  \underbrace{\bm{\lambda_{2}}\displaystyle\sum\limits_{\mathclap{j \in \mathcal{N}^{i}}} \min~ \left[\bm{0},\frac{\norm{\bm{q^{i}} - \bm{q^{j}}}^{2} - D^{2}}{\norm{\bm{q^{i}} - \bm{q^{j}}}^{2} - d^{2}}\right]}_{\text{repulsive term}}.
\end{equation} 
Here, $\bm{q^{i}_{\mathcal{T}}}$ is the target position of the astrobot $i$. $\mathcal{N}^{i}$ denotes the set of the neighboring astrobots of the astrobot $i$. The radius of the safety envelope around each astrobot is $D$. $d <D$ represents the safety region's radius of each astrobot. A safe control law can be immediately derived from the function above as below. 
\begin{equation}
\bm{u_{i}} := -k\nabla\psi(\bm{q^{i}})
\end{equation}
Here $k$ is a position step parameter which determines the intensity of each motion step. 

The control law above, interpreted as a velocity profile, successfully avoids collisions. However, it cannot resolve many deadlock cases in which two or more astrobots block each other's ways, turning into infinite oscillating motions. To handle deadlocks, a priority coordination mechanism is taken into account. This mechanism is implemented by a finite state machine (FSM) \cite{tao2018priority}. The resulting nonlinear hybrid controller is constituted by two decision layers. A low-level navigator, i.e., a DNF, first governs the coordination process until it faces a problematic deadlock case. Then, a high-level decision maker, say, an FSM, exploits the priority order of astrobots or their targets to manage the deadlock. This strategy improves the convergence rate of a typical astrobots swarm up to $\sim$\%85. Namely, this framework controls each astrobot in a selfish manner, say, once an astrobot reaches its target, it refuses any extra motion despite the potential case of occluding the paths of some of its neighbors to reach their targets. Put differently, the aforesaid formulation is not able to completely coordinate a general astrobots swarm associated with a particular observation in a guaranteed fashion. However, complete coordination is particularly interesting according to which the information throughput of an observation is maximized.
\subsubsection{Complete Safe Coordination}
\begin{figure*}
	\centering
	\hspace*{0mm}
	\begin{subfigure}[b]{\textwidth}
		\centering
		\includegraphics[scale=1]{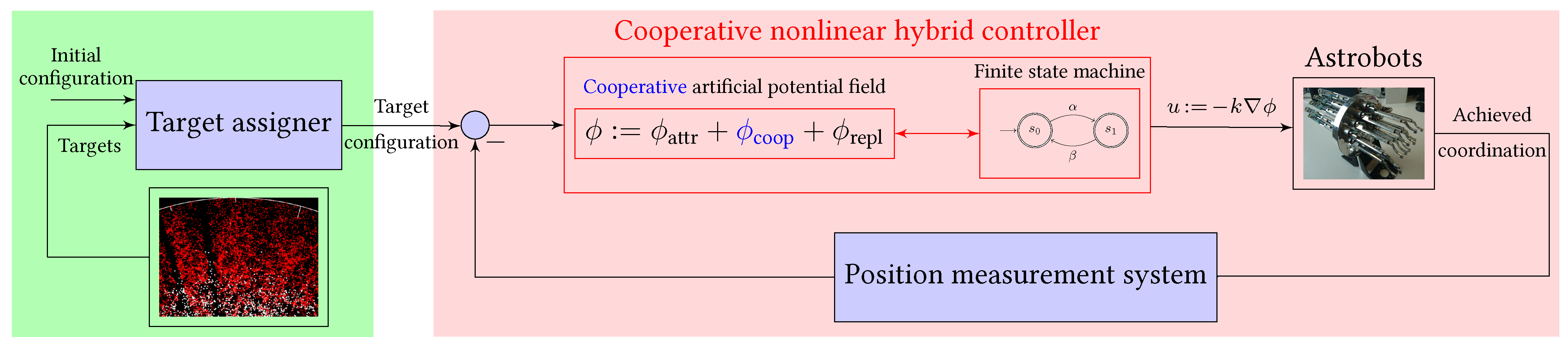}
		\caption{\label{fig:loop}}%
		\vspace{3mm}	
		\begin{minipage}[b]{.25\linewidth}
			\centering
			\subcaptionbox{\label{fig:1}}
			{\includegraphics[scale=0.12]{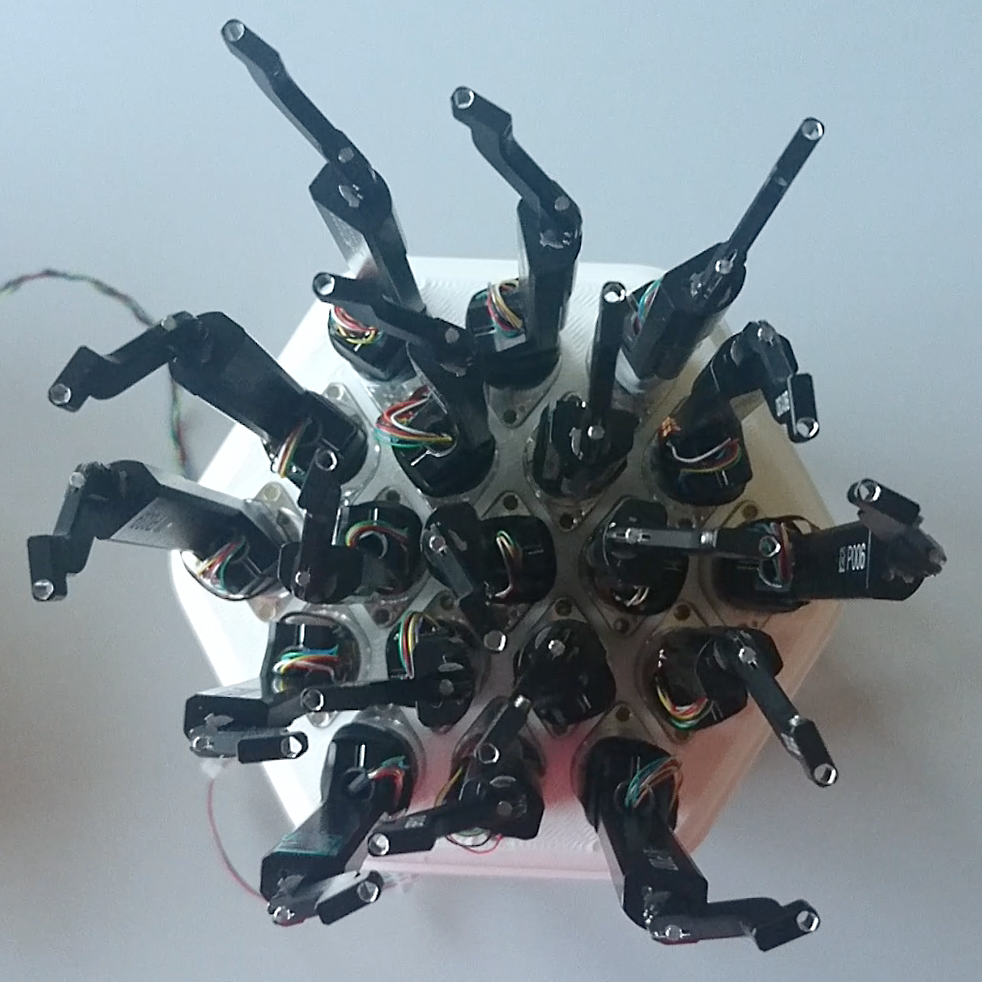}}
		\end{minipage}
		\begin{minipage}[b]{.25\linewidth}
			\centering
			\subcaptionbox{\label{fig:2}}
			{\includegraphics[scale=0.12]{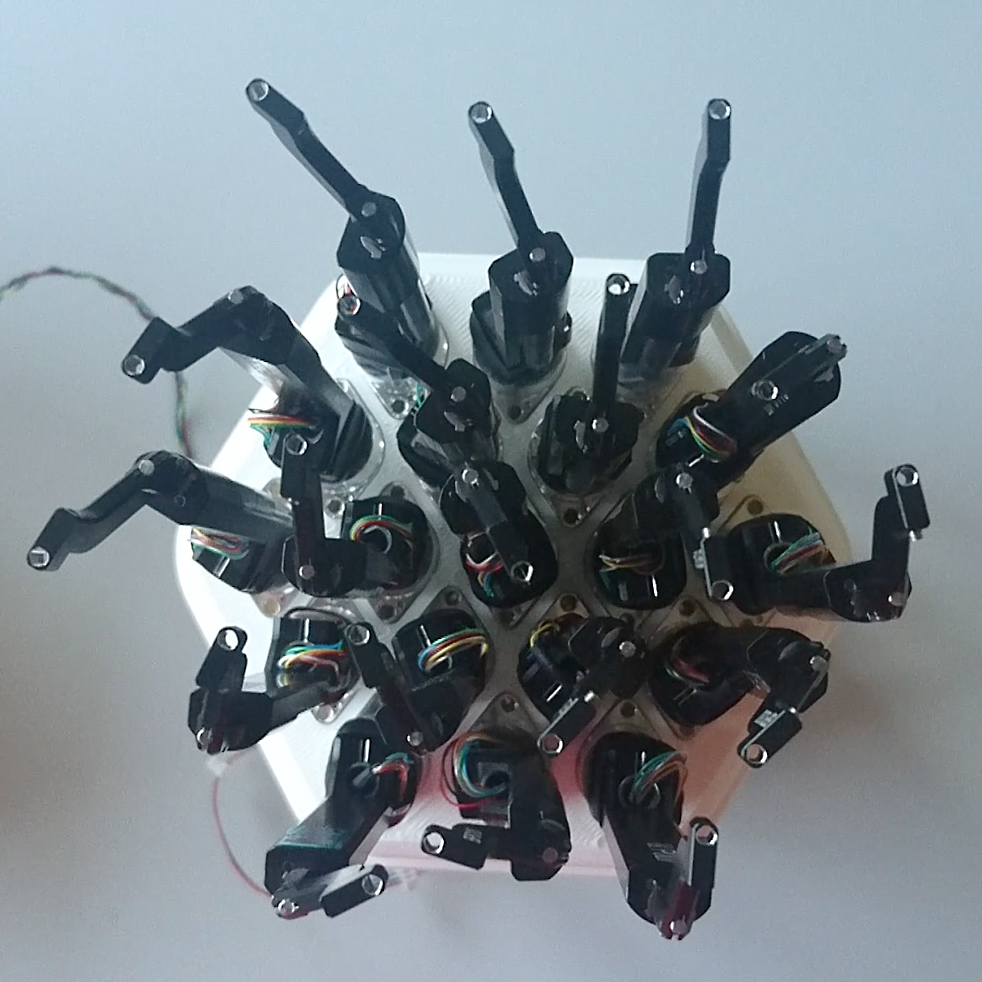}}
		\end{minipage}
		\begin{minipage}[b]{.25\linewidth}
			\centering
			\subcaptionbox{\label{fig:3}}
			{\includegraphics[scale=0.12]{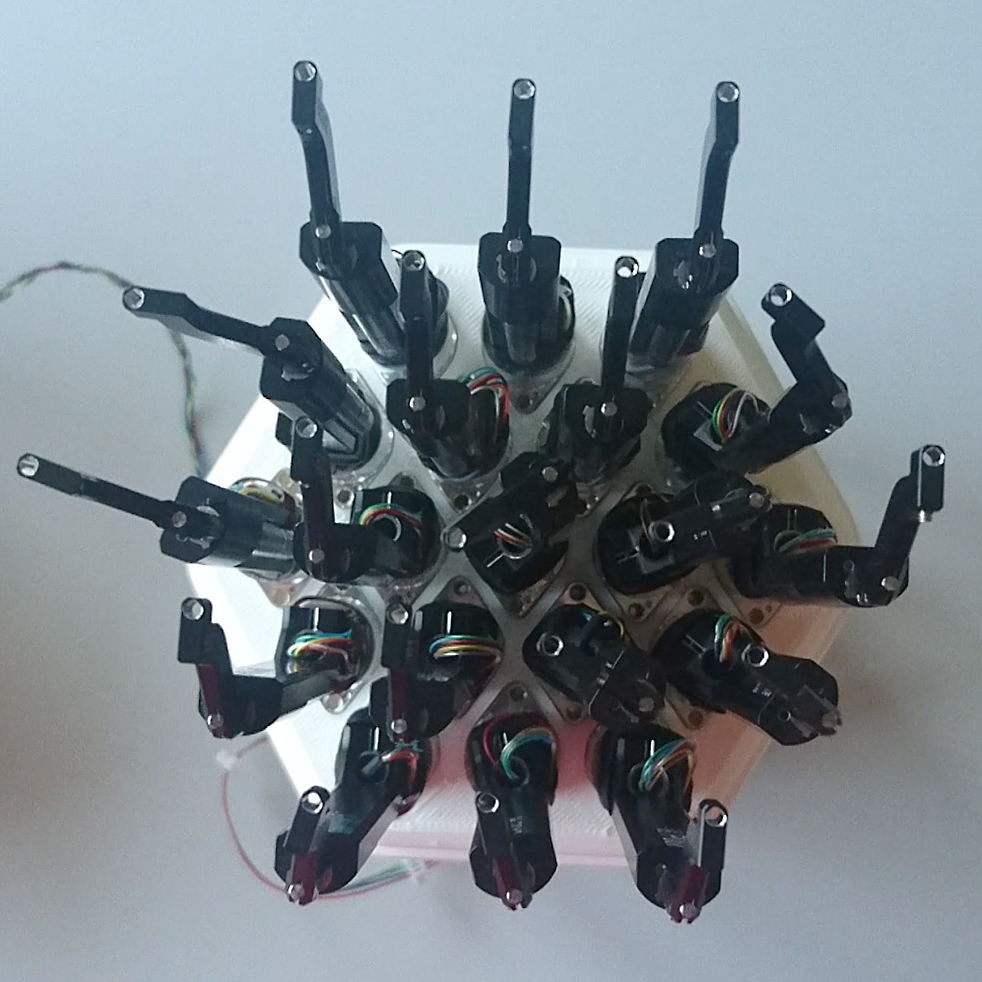}}
		\end{minipage}
		\begin{minipage}[b]{.25\linewidth}
			\centering
			\subcaptionbox{\label{fig:4}}
			{\includegraphics[scale=0.12]{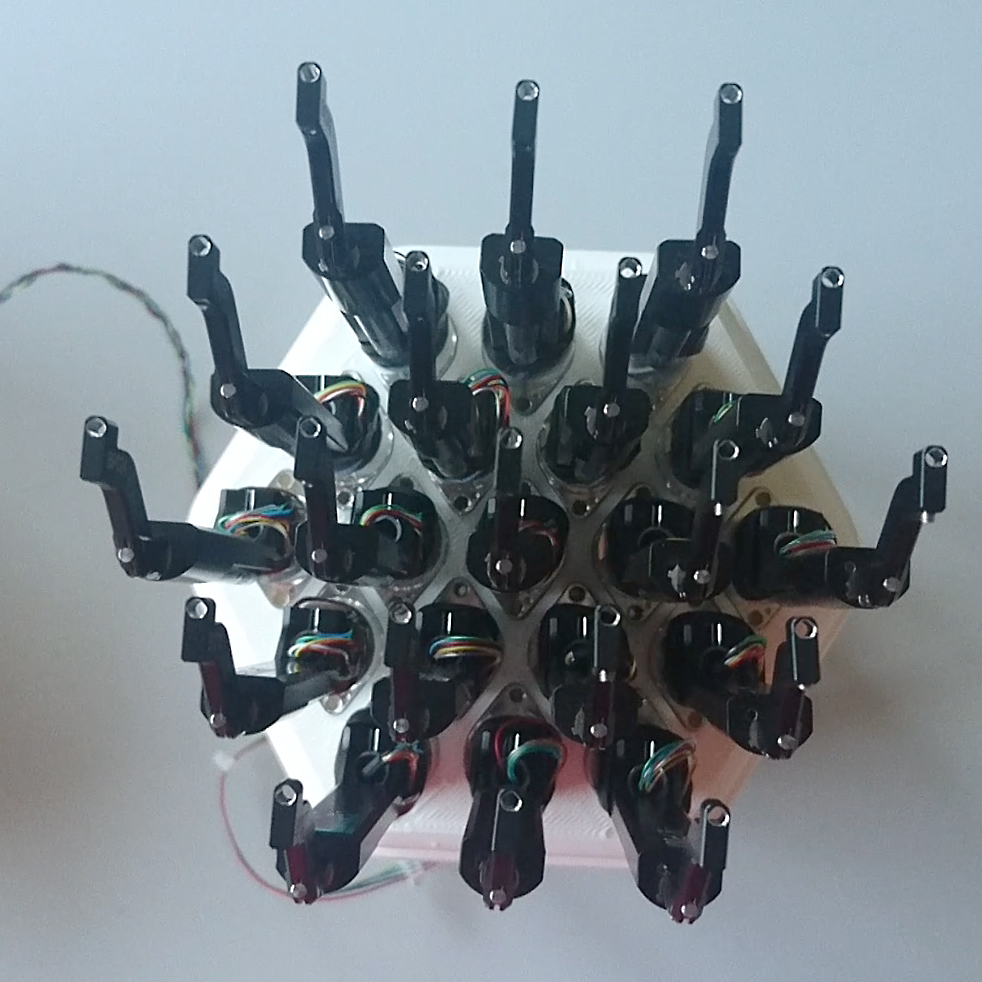}}
		\end{minipage}
		\end{subfigure}
	\caption{Completeness seeking using CNH control strategy. (a) Control architecture. \y{(b-e) A typical coordination process, governed by CNH, started from a random initial configuration to a synchronized one.}}
\end{figure*}
A DNF is unable to completely coordinate a swarm of astrobots. One may note that the coordination of astrobots may be better done in a cooperative, rather than a competitive, manner. In the competitive viewpoint, an astrobot does not care about the convergence of its neighboring counterparts. Thus, the convergence rate may be dramatically reduced if many astrobots in a locality need to cross an area corresponding to a particular already-converged astrobot. Alternatively, a cooperative perspective requires astrobots to not only seek their own convergences but also those of their neighbors. So, the overall completeness of the swarm is taken into account. For this purpose, the idea of \textit{cooperative artificial potential fields} (CAPF) \cite{macktoobian2019complete} is proposed. Each CAPF includes an extra cooperative attractive term as follows.
\begin{equation}
\label{eq:CAPF}
\phi(\bm{q^{i}}) := \underbrace{\vphantom{\bm{\lambda_{2}}\displaystyle\sum\limits_{j \in \mathcal{I}_{\mathcal{N}^{i}}\setminus \{i\}} \min [0,\frac{\norm{\bm{q^{i}} - \bm{q^{j}}}^{2} - D^{2}}{\norm{\bm{q^{i}} - \bm{q^{j}}}^{2} - d^{2}}}\bm{\lambda_{1}}\norm{\bm{q^{i}} - \bm{q^{i}_{\mathcal{T}}}}^{2}}_{\text{attractive term}} + \
\underbrace{\bm{\lambda_{2}}\displaystyle\sum\limits_{\mathclap{j \in \mathcal{N}^{i}}} \min~ \left[\bm{0},\frac{\norm{\bm{q^{i}} - \bm{q^{j}}}^{2} - D^{2}}{\norm{\bm{q^{i}} - \bm{q^{j}}}^{2} - d^{2}}\right]}_{\text{repulsive term}} + \underbrace{\bm{\lambda_{3}}\displaystyle\sum\limits_{\mathclap{j \in \mathcal{N}^{i}}}\norm{\bm{q^{j}} - \bm{q^{j}_{\mathcal{T}}}}^{2}}_{\text{cooperative attractive term}}
\end{equation}   	 
The cooperative term indeed injects extra dynamics to the velocity profile of its corresponding astrobot. So, if the astrobot is converged yet blocking some other peers' paths, it may temporarily leave its final spot to \z{provide} some free space for the blocked peers. This strategy successfully yields a completeness condition which can be locally evaluated based on the specifications of astrobots in various states, e.g., their initial and target spots. In particular, it is also formally shown that if all of the neighborhoods are completely coordinated, then the coordination of the whole astrobots swarm associated with those neighborhoods is globally complete \cite{macktoobian2019complete,macktoobian2020experimental}. Thus, the nonlinear hybrid control discussed in the previous section is substituted with the paradigm of \textit{cooperative nonlinear hybrid} (CNH) control as depicted in Figure \ref{fig:loop}. \y{Figures \ref{fig:1}-\ref{fig:4} illustrates the steps of a coordination process, governed by CNH scheme, in which an initial amorph configuration is coordinated to a synchronized one.} 

The CNH control scheme efficiently realizes complete coordination. However, it is computationally impossible to formally verify the safety and the correctness of the solutions obtained by it. Thus, one has to employ a set of simulations to check whether or not the generated signals indeed fulfill the requirements of their coordination process. Such numerical verification procedures are time-consuming and need excessive computational resources. Alternatively, supervisory control theory \cite{wonham2018supervisory} and discrete-event systems are legit options to yield verifiability. In this regard, a new derivation of the attraction field idea \cite{macktoobian2017automatic}, i.e., backtracking forcibility, can be effectively used to coordinate astrobots swarms, as well. In particular, \cite{macktoobian2019supervisory} established a verifiable supervisory coordination solution which determines whether or not a particular observation setup of astrobots can be completely coordinated. 
\begin{figure}
	\centering
	\includegraphics[scale=0.8]{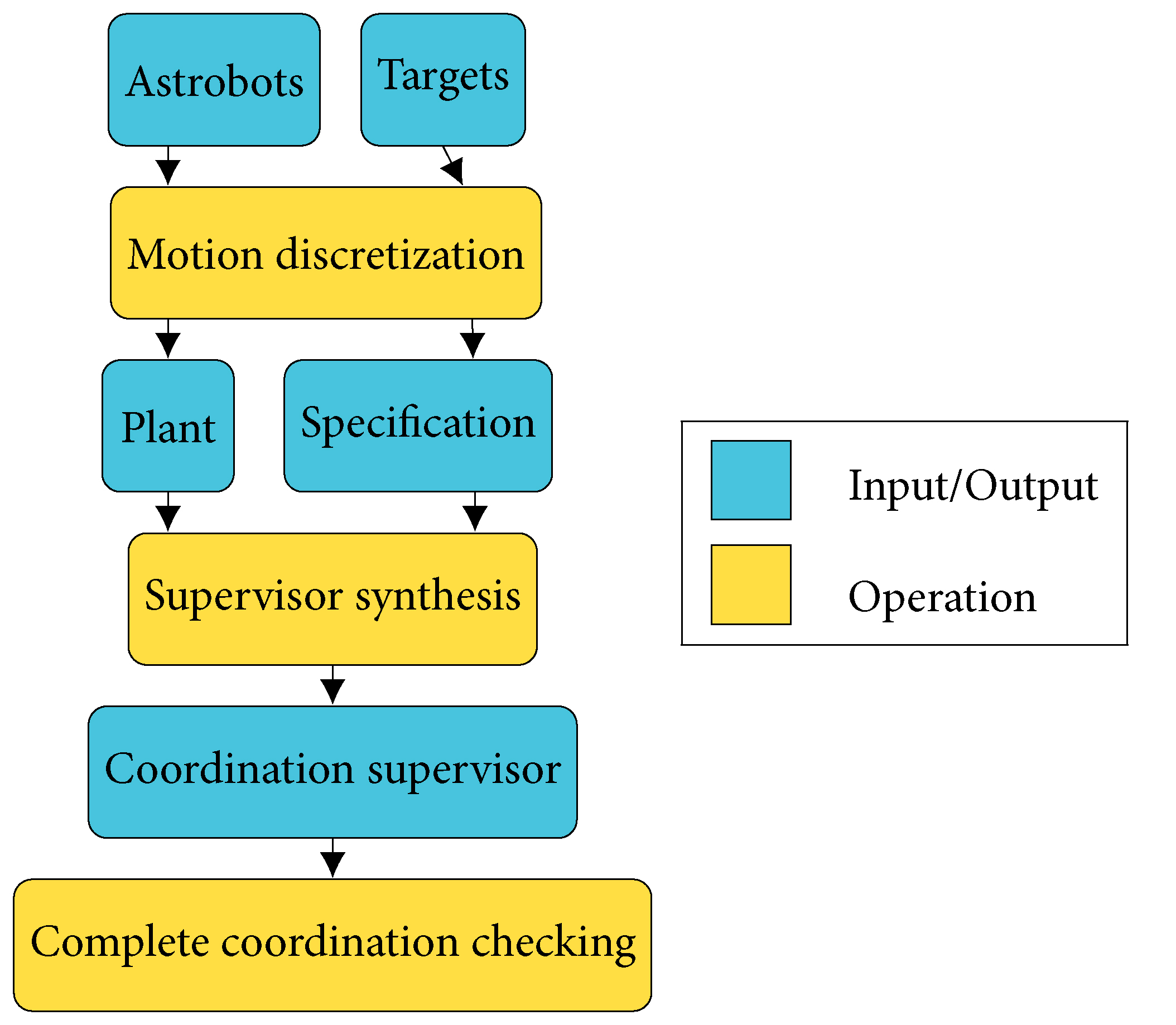}
	\caption{Completeness seeking via supervisory coordination}
	\label{fig:app}
\end{figure}

The supervisory coordination pursues the completeness feasibility checking as illustrated in Figure \ref{fig:app}. First, the continuous working space of astrobots are discretized to be embedded into discrete-event models. Then, each astrobot is encoded by a discrete-event model considering its particular characteristics such as its neighborhood map and its priority with respect to its neighboring astrobots. The desired behavior of each astrobot is also formulated in a discrete-event model as a specification. Namely, the cited behavior comprises various step-wise motions which lead to the target spot of the astrobot with respect to its initial point. Then, all of the astrobots' discrete-event models are synchronized to obtain an overall discrete-event plant model of the intended astrobots swarm. Then, a coordination supervisor is synthesized each of whose states represents a particular configuration of the swarm. In this formalism, a coordination is defined as any forcible path from one state to another. Thanks to the specific discrete-event formulations of astrobots and specifications, the coordination supervisor encompasses only one final state which constitutes the completely coordinated configuration of the swarm. Then given an initial state (configuration), the completeness seeking problem reduces to finding a forcible path from that state to the marked state (completely coordinated configuration) of the coordination supervisor.
\section{Trends in Future Research}
\subsection{Multi-Fiber Astrobot Design}
There are various types of fibers each of which can capture signals corresponding to different ranges of the electromagnetic spectrum. In particular, optical and near-infrared are extensively used in ground telescopes for spectroscopic purposes. Each type of electromagnetic radiation reveals unique information about the object which emanates it. Thus, spectroscopic surveys may be enriched if more than one type of signal can be simultaneously taken into account corresponding to each object of an observation. This requirement indeed leads to the emergence of many challenges in terms of astrobot design, collision avoidance, and spectrograph processing.

The ferrule of each state-of-the-art astrobot can only host one single fiber. So, multi-fiber ferrules have to be designed. The first challenge would be the relative arrangement of end-effectors so that all of them can observe a particular target. In this case, the kinematics of the current artifact, i.e., (\ref{eq:kin}), has to be revisited. It turns out that the control of each astrobot is more prone to nonlinearity of the base-to-end-effector transformations that, in turn, increase the approximation-based errors of an astrobots' control system. 

Multi-fiber astrobots may also be more subject to collision avoidance compared to mono-fiber ones. In particular, the ferrule of a multi-fiber astrobot is larger than that of a mono-fiber one. Each multi-fiber astrobot \z{may occupy} the working space of its neighboring peers more than what one observes in the mono-fiber localities. So, the definition of safety zone becomes critical, and the upper-bound of the motors' velocities have to decrease not to jeopardize the safety of neighboring astrobots.
\subsection{Distributed Target-to-Fiber Assignment}
The next-generation giant surveys \cite{schlegel2019astro2020} include many thousands of targets and astrobots, say, $\sim$20,000. In particular, optimal target assignment method may not be effective enough to be applied to \x{surveys of this size}. Namely, optimal target assignment takes minimum coordination and maximum distribution requirements into account to perform the assignments. But in the case of extremely complex astrobots swarm of giant surveys, the cost function of the method may need to be revised to consider additional requirements. For example, if one or more astrobots are malfunctioned for any reason, then the coordination process may be adversely impacted in terms of the noticeable convergence rate loss of functional astrobots. This issue may be (at least partially) resolved by assessing other assignment options to minimize the negative influence of the cited malfunctions. To this aim, a key question is whether target assignment can be revised to be performed in a real-time manner. One may even imagine the development of adjustable focal planes which can be moved according to some degrees of freedom. So, in the case of such undesirable occurrences, trivial adjustment of focal planes may be helpful.
\subsection{Miniaturization and Density-Safety Trade-Off}
\begin{figure}
	\centering
	\includegraphics[scale=0.6]{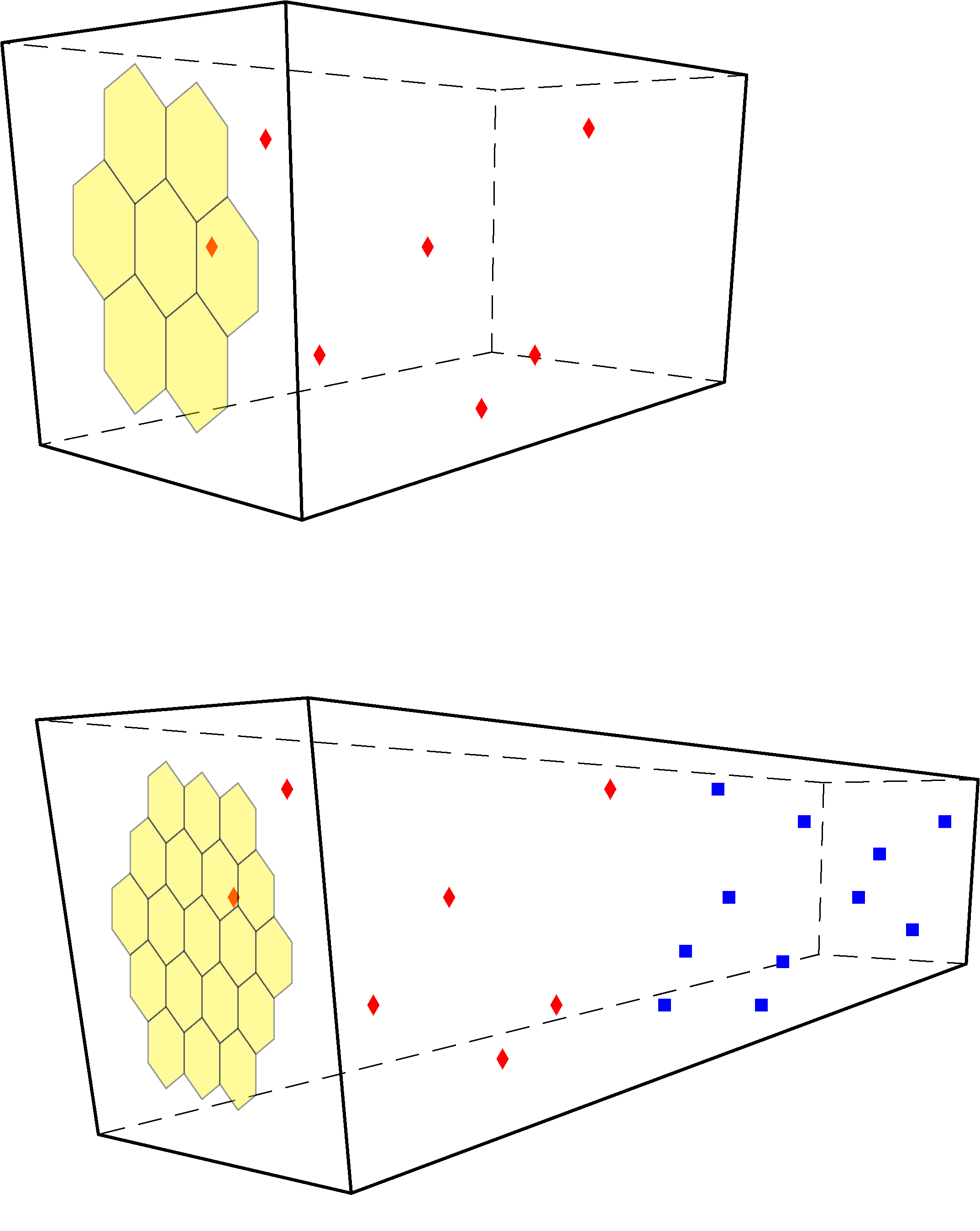}
	\caption{The increment of observation depth by increasing the density of astrobots with respect to a given field of view}
	\label{fig:FoV}
\end{figure}
The more astrobots are available at a focal plane, the more targets can be observed. Put differently, the more dense an astrobots swarm is, the deeper observations can be taken into account without any change in the field of view. As depicted in Figure \ref{fig:FoV}, the upper sketch represents a set of astrobots (drawn in yellow) covering a particular area of a hypothetical focal plane. So, the red targets can be observed using such a setup. In the lower sketch, the observational field of view is not changed, but focal plane is filled by smaller astrobots. In this case, more targets (red and blue ones) can be observed compared to the upper case. In other words, miniaturization of astrobots, given a fixed area of focal focal plane, expands the set of targets which can be observed, thereby the depth of their resulted survey. So, further miniaturization of astrobots strengthens the information reflected by surveys.

The effective reduction of astrobot size is contingent on resolving many mechanical difficulties. For example, \x{not only is there a need for smaller bearings, but efficient interfaces between them and motors must be assembled}. The shaft stiffness is a determinant factor to retain the nominal functional accuracy of an astrobot. A rigorous analysis is then required to compensate any stiffness loss corresponding to the shaft of an astrobot in the course of its extra miniaturization. The fiber design has to be revised through scaling astrobots down. Namely, the fiber rigidity does not scale with the astrobot's size. So, the negative impacts of miniaturization on the fiber rigidity need to be thoroughly counteracted.
\subsection{Efficient Coordination}
\subsubsection{Finite State Machine Reduction}
The hybrid aspect of the CNH controller clearly states the participation of a decision-making FSM in the controller formulation. As we already asserted, this FSM is taken into account as a utility to cover the deadlock situations which cannot be resolved by the planned CAPF. Despite the usefulness of the FSM, its presence adds another nonlinear layer to the overall controller architecture. This layer indeed hides the fairly intuitive mapping which corresponds a CAPF to its generated control law. We already noted that there is no formal verification method to assess the credibility of the generated control signals. Thereby, the more layers a controller has, the more simulation scenarios have to be run to assure the controller's fine-tuned behavior. A key question is how the CAPF formulation (\ref{eq:CAPF}) can be revised so that it can also cover the functionality of the FSM.

The improvement of the verifiable supervisory solution \cite{macktoobian2019supervisory} may also be the subject of active research. For instance, the current scheme formulates the discrete-event model of an astrobot as a one-degree-of-freedom artifact. This assumption implies that each astrobot has only one arm whose length is the overall length of the arms of an actual astrobot. The size of the discrete-event model of such a reduced astrobot is efficiently small to make synthesis computations fairly tractable. However, the elimination of one degree of freedom in fact neglects many potential coordination solutions which require the independent motions of both arms.
\subsubsection{Optimal Coordination}
The supervisory coordination strategy clearly exhibited how an astrobots coordination problem can be discretized with respect to the step-wise motions its astrobot's arms. We also studied the critical role of the forcible backtracking to find complete solutions. In view of the supervisory coordination, given a particular astrobot, all of the event paths which fulfill both the safety and the completeness requirements are the same. But one is logically more interested in the shorter solutions whose execution times are faster. Thus, optimized trajectory planning through the discrete motion space of astrobots is of utmost interest. The temporal efficiency of a path or its likelihood not to end up with a deadlock scenario can be reflected by a weight. Then, the whole graph of trajectories with respect to an initial configuration may be traversed to reach a target configuration using dynamic programming. The challenge would be encoding the whole motion space of an astrobots swarm to an efficient graph structure.

Astrobots and focal plane comprises many numerical parameters which specify their mechanical structures. These parameters are kept in particular configuration files which are used in the artificial potential computations in CNH controllers. The impact of each parameter on the cited potential calculations is not obvious. So, if one changes one parameter, there is not a direct way, except intensive simulations, to observe how the change influences a coordination. Accordingly, each parameter may be taken as a constraint, or objective, into account in view of coordination. Then, the constraints can be directly applied to a model predictive controller.
\subsubsection{Coordination Prediction}
Both of the CNH-based and supervisory coordination require intensive computations to solve a coordination scenario. Given a setup, there is necessarily not a complete solution corresponding to it. Even in the case of the acceptance of partial convergences, a final result may not fulfill the minimum desired convergence rate of an observation. In such situations, intensive coordination computations are done in a blind manner without yielding what one looks for. One may not spend computational resources on the setup scenarios in which the chance of obtaining their coordination objectives is not considerable. Thus, the idea of coordination prediction comes to mind \cite{macktoobian2020data}. 

\begin{figure}
	\centering
	\includegraphics[scale=0.8]{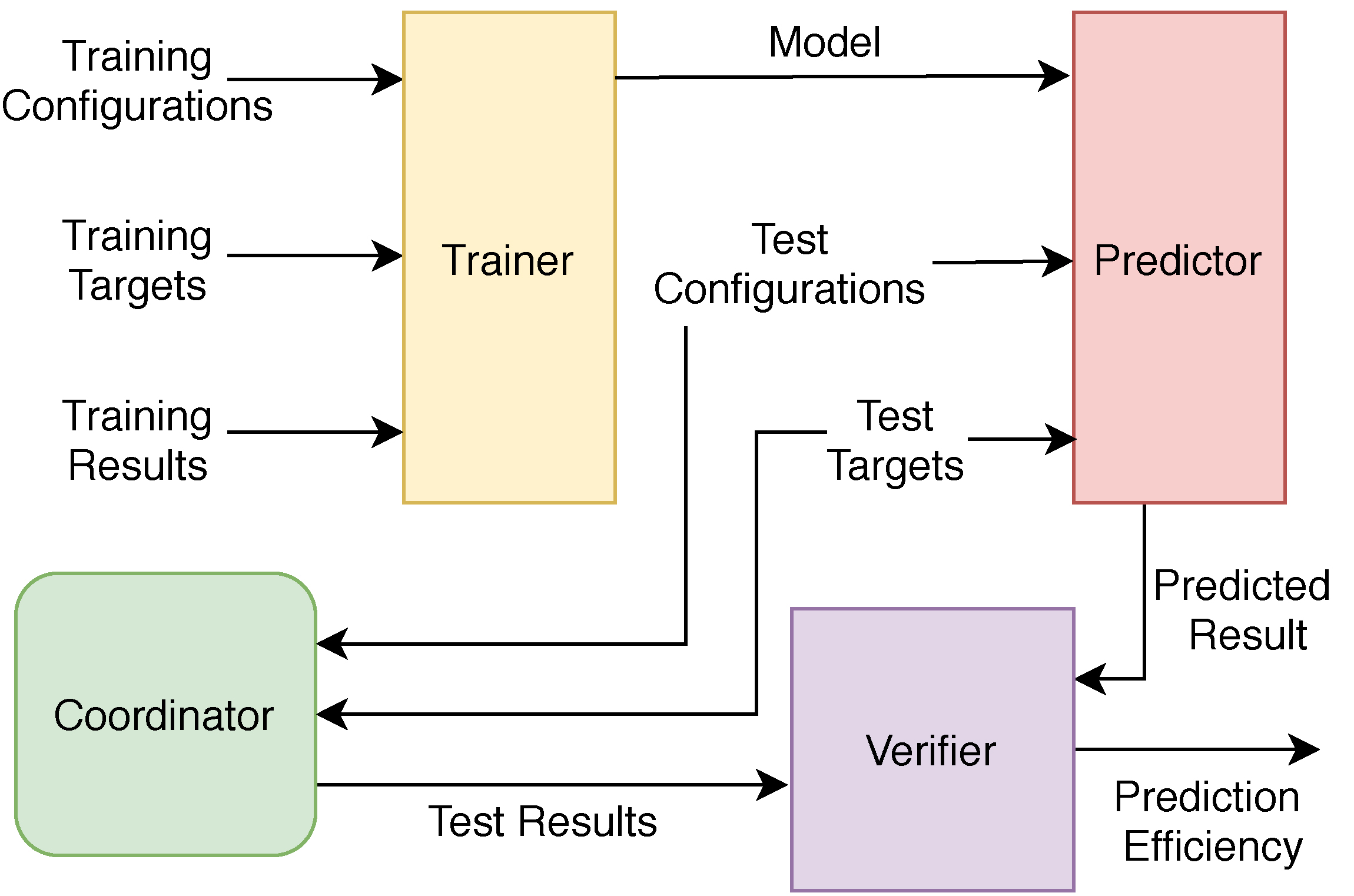}
	\caption{The schematic flow of a potential coordination prediction process}
	\label{fig:pred}
\end{figure}

To train a predictor model (see, Figure \ref{fig:pred}), we first feed a class of training astrobots' configurations, their corresponding set of targets, and the results of their coordination into a machine learning algorithm. Coordination prediction may be pursued as a classification problem. In this regard, the training results set only includes true/false entries representing complete and partial coordination, respectively. Alternatively, the results may be a set of probabilities predicting the likelihood of completeness. The test configuration used in the model generation and the test intended targets sets are fed into a coordinator governed by a CNH controller. The test results obtaining from the coordinator are compared to the predicted results to check the success rate of the model.
\subsection{New Astrobotics Applications}
\z{Astrobots have been greatly contributing to astrophysical studies and observational cosmology. However, \x{astrobot} swarms are versatile enough to be used in \x{a variety of} space-related applications. One example is the detection of space debris to support space debris removal missions \cite{nishida2009space}. In particular, there \x{is} a huge amount of space debris around the Earth, \x{e}specially in the LEO\footnote{LEO refers to the low-Earth orbit which generally corresponds to the Earth-centered orbits at the altitude range of 200-1600 km.} which needs to be mitigated for various reasons. First, these objects are revolving around the Earth with very high velocities. So, \x{if they impact another satellite, the satellite may obtain substantial damage and cease to function}. On the other hand, many of these objects are huge enough to be considered as l\x{arge enough to pose a significant threat to Earth}. Namely, \x{there are non-operational satellites in low Earth orbits}. \x{These satellites may enter the Earth atmosphere, where they may completely burn up in the atmosphere or may hit the Earth's surface}. So to remove such hazards, these satellites need to be safely de-orbited. Additionally, some asteroids and comets which orbiting the Earth in LEO range include valuable minerals which are no longer frequently found on the Earth. Thus, asteroid engineering missions intend to capture and excavate such asteroids to obtain those minerals \cite{anthony2018asteroid}. To mitigate a piece of space debris, debris collector satellites have been launched \cite{pirat2017mission} whose sensor range to detect debris in their proximity does not exceeds some kilometers. Moreover, space debris detection using GPS satellites is not economical and even generally possible. The reason is that debris collection requires the spectrum analysis of particular electromagnetic signals which is beyond the capabilities of many satellites. In this regard, astrobots have already shown \x{potential} to contribute to space debris tracking missions \cite{macktoobian2019heterogeneous}. In particular, we explained the existence of the fibers which capture near-infrared rays. The objects rotate in LEO range with very high velocities. So, the friction of these objects with the thin atmosphere increases their surface temperature. Thus, they radiate near-infrared waves which are detectable by near-infrared fibers \cite{birlan2020first}. One may note that a relatively large focal plane equipped with these fibers can cover a vast area of the LEO-range space compared to what a single in-orbit satellite can do. This application simply signifies that the investment on astrobots and spectrograph-equipped ground telescopes can potentially \x{lead} to their extensive \x{utilization} in many science and technology disciplines. In summary, we argue that more applications of astrobotics may gradually emerge to contribute to scientific discoveries and technological developments.}

\z{Satellites observatories have already contributed to the generation of spectroscopic surveys \cite{brammer20123d}. However, because of the limited capacities of these \x{systems}, their surveys are not comparable to those massive ones of ground telescopes in view of resolution. The future miniaturization of astrobots may pave the way for generating multi-fiber spectroscopic surveys in space.} 
\section{Conclusion}
Dark \z{matter} studies have been fascinating enough to astrophysicists to seek the map of the whole observable universe. The cosmological spectroscopy benefits from a plethora of optical fibers whose configurations toward sky have to be revisited from one observation to another. The generation of the desired map needs massive numbers of optical fibers to be automatically coordinated. Thus, astrobotics field is recently emerged whose \x{design and operation} open a wide range of challenging theoretical and technological problems. In particular, we explained how astrobots constitute swarm robots which exhibit unique issues stemming from their specific requirements. Despite the fairly simple structure of a single astrobot, we addressed the extremely challenging nature of coordinating their swarms so that both collision freeness and completeness requirements are fulfilled. 

The target assignment problem was introduced noting the importance of this problem and its influence on the quality of its subsequent coordination phase. The efficiency of optimal target assignment was asserted in the case of massive spectroscopic surveys. However, The key open question is how to localize optimizations in small neighborhoods of astrobots swarms in the course of assignment processes. Furthermore, the smaller astrobots become, the more artifacts can be placed in focal plane areas, thereby, the higher information throughput of surveys are. In particular, stiffness and rigidity of astrobots need to be compensated in smaller scales. The development of MEMS-based motors and their bearing is also a must. We also reviewed the two leading candidate methods to safely and completely coordinate astrobots swarms, say, CNH control and supervisory control. The CNH control strategy seeks completeness a whole swarm via the accumulation of locally complete partitions of that swarm. We introduced two potential options to optimally control coordination, say, dynamic programming and model predictive control. We outlined how machine learning ideas may be used to predict completeness of coordination. \z{The application of astrobots in tracking space debris in LEO range also manifested the prospect of further expansions of their applications in space and astronomical studies.}

\nocite{*}
\bibliographystyle{IEEEtran}
\bibliography{references}{}

\end{document}